\documentclass[12pt, letterpaper]{article}
\usepackage{geometry}
\usepackage{amsmath}
\usepackage{graphicx}
\usepackage{float}
\usepackage{siunitx}
\usepackage{empheq}
\usepackage{subcaption}
\usepackage{amssymb}
\usepackage{cite}
\usepackage{stackengine}
\newgeometry{vmargin={18mm}, hmargin={18mm,18mm}}
\usepackage{mathtools}
\usepackage{ragged2e}

\DeclarePairedDelimiterX\braket[2]{\langle}{\rangle}{#1\,\delimsize\vert\,\mathopen{}#2}

\providecommand{\keywords}[1]
{
  \small	
  \textbf{\textit{Keywords---}} #1
}
\begin{document}
\title{Supersymmetric pairing of Lambert $W$–kink nerve impulses}
\author{M. F. De la Rosa-López $^{1}\footnote{mrosal001@alumno.uaemex.mx}$, D.M. Galv\'an-Arellano$ ^{2}\footnote{dulce\_galvan@my.uvm.edu.mx}$ ,  J. L. Larios-Ferrer$^{3}\footnote{leonel.larios@upenergia.edu.mx}$,  \\ V. A. Mendoza-Millán $^{1}\footnote{vmendozam002@alumno.uaemex.mx}$, O. Pav\'on-Torres$^{4}\footnote{omar.pavon@cinvestav.mx \text{(corresponding author)}}$}
\date{%
\small{$^{1}$ Facultad de Ciencias, Universidad Autónoma del Estado de México, Toluca 50200, M\'exico\\
\small{$^{2}$ Universidad del Valle de México, Campus Toluca, Metepec 52164, Mexico\\}
\small{$^{3}$ Universidad Polit\'ecnica de la Energ\'ia, 42820, Tula de Allende, Hidalgo, México\\}
\small{$^{4}$ Physics Department, Cinvestav, POB 14-740, 07000 M\'exico City, M\'exico}\\ 
}}
\maketitle

\begin{abstract}
Nerve impulses can be modelled as electromechanical density waves within the improved Heimburg–Jackson model.  The inclusion of higher-order polynomial nonlinearities leads to a generalized Boussinesq equation with third and fourth order nonlinearities that, under a traveling-wave reduction, reduces to a Liénard-type equation. Applying a factorization method yields exact Lambert $W$–kink soliton solutions that represent localized nonlinear density waves near the membrane melting transition. Beyond providing exact solutions, the factorization uncovers an underlying supersymmetric structure. The associated operators satisfy algebraic relations analogous to those of supersymmetric quantum mechanics, thereby enabling the construction of a partner soliton. This supersymmetric pairing establishes a novel and previously unexplored connection between nonlinear electromechanical wave propagation in biological membranes and supersymmetric quantum-mechanical methods. The resulting framework offers a theoretical foundation for analysing mechanically induced perturbations and their nonlinear propagation in nerve membranes, with potential implications for understanding the biomechanical mechanisms underlying traumatic brain injury.
\end{abstract}

\! \! \! \! \keywords{Factorization method, Supersymmetric pairing, Lambert $W$-Kink solitons.}

\begin{justify}
\section{Introduction}

Signal propagation in excitable cells is governed not only by electrical activity but also by intrinsic physical properties such as axon diameter, myelination, elasticity, and membrane structure, which play a fundamental role in nerve impulse transmission \cite{tejocote1, tejocote2, tejocote3, tejocote4, tejocote5, tejocote6, tejocote7, tejocote07, tejocote007, tejocote0007}. To account for these features and achieve a more comprehensive description of nerve signal propagation, several theoretical frameworks beyond the classical Hodgkin–Huxley model have been proposed, in which nerve impulses are interpreted as electromechanical waves. These approaches can be broadly classified into four categories: (i) classical electrophysiological models focusing on electrical dynamics \cite{tejocote8, tejocote9, tejocote10, tejocote11}; (ii) thermodynamic models addressing energy exchange and heat-related effects \cite{tejocote12, tejocote13}; (iii) mechanical models incorporating membrane elasticity and density variations \cite{tejocote14, tejocote15, tejocote16}; and (iv) hybrid electromechanical models that explicitly couple electrical and mechanical degrees of freedom \cite{tejocote17, tejocote18}. More recently, fractal-based approaches have been introduced to capture scale-invariant behaviour and memory effects in neuronal systems, extending continuum descriptions of nerve dynamics \cite{tejocote19, tejocote20, tejocote020}. Collectively, these frameworks have been applied to phenomena such as head-on pulse collisions, anesthesia, mechanosensory responses, phase transitions, refractory effects, and adiabatic signal propagation \cite{tejocote21, tejocote22, tejocote23, tejocote24}.

In particular, the Heimburg–Jackson (HJ) model provides a unified thermodynamic and mechanical framework for describing nerve signal propagation along the axon. This model has been extensively studied, with significant effort devoted to deriving exact travelling-wave solutions using quasi-analytical methods \cite{tejocote25, tejocote26, tejocote27, tejocote28, tejocote29, tejocote30, tejocote31, tejocote32}. Within this framework, phase transitions in lipid membranes triggered by action potential propagation have been widely investigated \cite{tejocote33}. The associated density waves are commonly described as hyperbolic kink solutions arising from Boussinesq-type equations in both the standard and improved HJ models \cite{tejocote34, tejocote35, tejocote36, tejocote37}. More recently, it has been shown that introducing strong nonlinearities in extended versions of the improved HJ model deforms these hyperbolic kink into Lambert $W$–type kink solitons \cite{tejocote38}. 

In general, such nonlinear extensions render the improved HJ model non-integrable. The resulting governing equation can be reduced to a Li\'enard-type equation with constant damping, a structure that plays a central role in its analytical treatment. Integrability properties of Li\'enard-type systems have long been recognized through their connections with scalar field theories such as the $\phi^{4}$ and $\phi^{6}$ models \cite{tejocoteprepre39}. In particular, polynomial nonlinearities and BPS-type reductions allow the second-order dynamics to be mapped into first-order equations derived from energy minimization principles \cite{tejocotepre39}. Alternatively, factorization methods provide a direct algebraic alternative to obtain reduced first-order systems, offering an efficient framework for solving Li\'enard-type equations \cite{tejocote39, tejocote40}. This procedure reduces the Li\'enard equation to an Abel equation of the first kind with constant coefficients, thereby enabling the explicit construction of exact solutions. In this context, factorization under constant damping into a product of two non-commuting first-order differential operators becomes particularly relevant, since the ordering of the operators directly affects the resulting reduced equations and their solvability.

A further consequence of this non-commutative structure is the emergence of paired solution sets associated with reversed operator orderings. In this framework, the Li\'enard equation admits two related factorized forms, giving rise to partner equations connected through an underlying algebraic transformation \cite{tejocote41}. This correspondence induces a structured pairing of solutions that share global dynamical properties, such as propagation velocity, while differing in their profiles and effective parameters. From this perspective, the pairing reflects an algebraic organization of the solution space generated by the nonlinear factorization scheme. Physically, such paired solutions may be interpreted as distinct electromechanical states of the membrane related through effective transformations induced by external perturbations, while the damping parameter fixes the admissible dynamical regime.

Motivated by these considerations, we show that the non-integrable structure of the Li\'enard equation, under suitable constraints on the nonlinear elastic coefficients, gives rise to Lambert $W$–type solitons whose multivalued analytic structure is consistent with the breakdown of the Painlevé property. Furthermore, we demonstrate that this structure admits a supersymmetric pairing through factorization, linking the analytic properties of the solutions with their underlying algebraic organization. The remainder of this paper is organized as follows. Section 2 outlines the improved HJ model with higher-order nonlinear terms and its reduction to a Li\'enard equation, together with a Painlevé integrability analysis. Section 3 presents the implementation of the factorization method and the hierarchy of factorization orders used to construct Lambert $W$–kink solitons and their supersymmetric partners, along with possible biological interpretations related to pathological damage and specially traumatic brain injury. Finally, Section 4 summarizes our main results and concluding remarks.

\section{Thermodynamic soliton theory of the nerve impulses}
Based on the thermodynamic behaviour associated with phase transitions in lipid bilayers and biological membranes, T. Heimburg and A. D. Jackson \cite{tejocote12}, together with S. T. Andersen et al. \cite{tejocote13}, proposed a model in which the nerve impulse is described as a nonlinear mechanical density wave propagating along a cylindrical biomembrane, as schematically illustrated in Fig. \ref{ig1}. 
\begin{figure}[ht]
\includegraphics[width=0.8\textwidth]{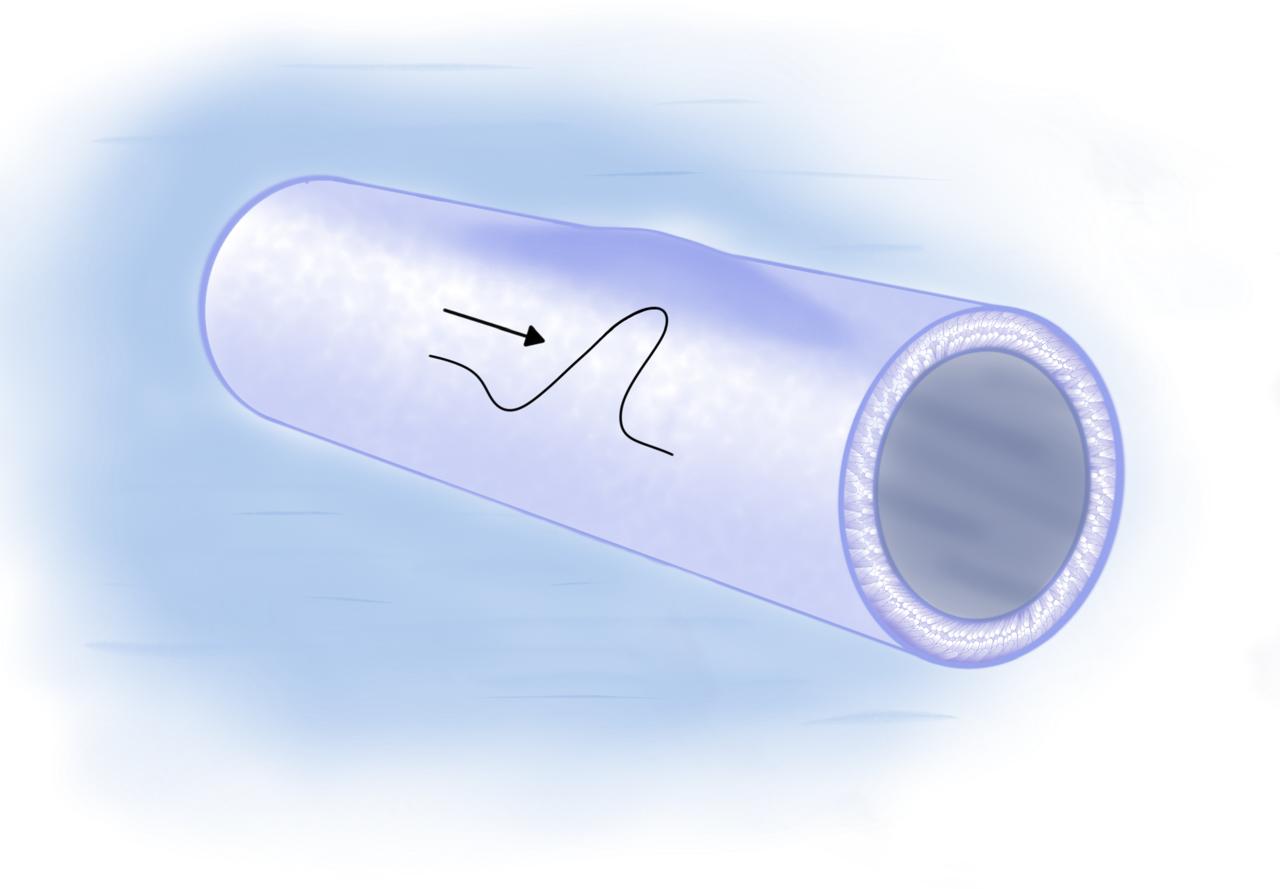}
\centering
\caption{Schematic representation of a cylindrical biomembrane and action potential produced by compression.}\label{ig1}
\end{figure}
Within this framework, the essential mechanism arises from the coupling between mechanical compression and membrane thermodynamics: local variations in density drive a reversible phase transition between the disordered liquid phase and the ordered gel phase. Consequently, the action potential can be interpreted as a localized compression pulse intrinsically linked to this phase transition. Moreover, due to the thermodynamic reversibility of the process, the inverse scenario is also admissible: local cooling may induce a transition toward the ordered phase, thereby generating a propagating density perturbation \cite{tejocote13}. 

In the present work, and in order to achieve our main objective, namely, to obtain and interpret the supersymmetric pairing of Lambert $W$–kink solitons, we adopt the following extended version of the HJ model \cite{tejocote42}. To this end, let $u=\rho^{A}-\rho_{0}^{A}$ denote the longitudinal density change, defined as the difference between the lateral mass density of the membrane $\rho ^{A}$ and  its empirical equilibrium value $\rho_{0}^{A}$. The governing equation is given by:
\begin{equation}
\dfrac{\partial^{2}u}{\partial {t} ^{2}}=\dfrac{\partial }{\partial x}\left(\left[c_{0}^{2}+\alpha u+\beta u^{2}+\epsilon u^{3}+\lambda u^{4}\right]\dfrac{\partial u}{\partial x}\right)-h_{1} \dfrac{\partial^{4}u}{\partial x ^{4}}+h_{2}\dfrac{\partial^{4}u}{\partial x^{2}\partial t ^{2}}+\mu \dfrac{\partial^{2}}{\partial x^{2}}\left(\dfrac{\partial u}{\partial t}\right). \label{ferrer1}
\end{equation}
The model is grounded on two key physical assumptions: (i) the existence of a phase transition in lipid bilayers, and (ii) the analogy between nerve impulse propagation and sound waves in a compressible membrane. The latter is evident from Eq. (\ref{ferrer1}), which reduces to a linear wave equation upon neglecting the terms proportional to $\alpha$, $\beta$, $\epsilon$, $\lambda$, $h_{1}$, $h_{2}$ and $\mu$. In the present limit of Eq. (\ref{ferrer1}), the propagation velocity $c$ is directly linked to the lateral compressibility, consistent with the original HJ hypothesis. Specifically, the sound speed in the fluid phase of the membrane is given by $c_{0}=1/\sqrt{K_{s} ^{A}\rho_{0}^{A}}$, where $K_{s} ^{A}$ is the lateral compressibility. The nonlinear elastic coefficients $\alpha$, $\beta$, $\epsilon$, and $\lambda$ are empirically determined parameters that encode distinct mechanical contributions, including lateral compressibility and membrane stretching, lipid rarefaction, compositional heterogeneity arising from lipid–protein interactions, and large-scale deformations induced by compressive and tensile stresses. The term proportional to $h_{1}$ represents the intrinsic elastic response of the biomembrane, whereas the $h_{2}$ term accounts for the inertial effects associated with lipid motion. Its inclusion promotes the HJ equation to a double-dispersion model, thereby mitigating the instabilities that typically emerge in formulations involving only spatial dispersion. Finally, the term proportional to $\mu$ models viscous dissipation due to the surrounding axoplasmic fluid. 

\subsection{Density wave equation}
In order to study the supersymmetric pairing of Lambert $W$-kink soliton solutions, we consider the following re-parametrization: 
\begin{equation}
z=\dfrac{u}{\rho_{0} ^{A}}, \quad  \zeta=\dfrac{c_{0}x}{\sqrt{h_{1}}}, \quad \text{and} \quad \tilde{t}=\dfrac{c_{0} ^{2}t}{\sqrt{h_{1}}},
\end{equation}
together with
\begin{equation*}
p=\dfrac{\alpha \rho_{0}^{A}}{c_{0}^{2}}; \quad q=\dfrac{\beta (\rho_{0} ^{A})^{2}}{c_{0} ^{2}}; \quad r=\dfrac{\epsilon (\rho_{0} ^{A}) ^{3}}{c_{0} ^{2}}; \quad s=\dfrac{\lambda (\rho_{0} ^{A}) ^{4} }{c_{0}^{2}}; 
\end{equation*}
\begin{equation}
\gamma=\dfrac{\mu}{\sqrt{h_{1}}} \quad \text{and} \quad \delta=\dfrac{h_{2}c_{0} ^{2}}{h_{1}}, 
\end{equation}
which leads to the following dimensionless form of Eq. (\ref{ferrer1})
\begin{equation}
\dfrac{\partial ^{2}z}{\partial \tilde{t} ^{2}}=\dfrac{\partial }{\partial \zeta}\left(\left[1+pz+qz^{2}+rz^{3}+sz ^{4}\right]\dfrac{\partial z}{\partial \zeta}\right)-\dfrac{\partial ^{4}z}{\partial \zeta^{4}}+\delta \dfrac{\partial ^{4}z}{\partial \zeta ^{2} \partial \tilde{t}^{2}}+\gamma \dfrac{\partial ^{3}z}{\partial \zeta^{2}\partial \tilde{t}}. \label{badillo1}
\end{equation}
To find travelling wave solutions of Eq. (\ref{badillo1}), we consider $z(\xi)$ with $\xi=k\zeta-v\tilde{t}$, where $k$ and $v$ are real constants. After two successive integrations, followed by an appropriate rescaling and setting the integration constants to zero without loss of generality, we obtain 
\begin{equation}
\dfrac{d ^{2}y}{d\xi ^{2}}+\tilde{\gamma}\dfrac{dy}{d\xi}-a_{1}y+a_{2}y^{2}-a_{3}y ^{3}-a_{4}y ^{4}-y ^{5}=0, \label{ferrer2}
\end{equation}
where the rescaled dependent variable is given by
\begin{equation}
z=\dfrac{1}{\tilde{s}^{1/4}}y.
\end{equation}
Additionally, the re-parameterized and rescaled constants $\tilde{\gamma}$, $\tilde{s}$ and  $a_{i}$ ($i=1, ..., 4$) are defined by
\begin{align}
\tilde{\gamma}=\dfrac{v \gamma}{\Lambda}; \quad  a_{1}=\dfrac{k ^{2}-v^{2}}{ \Lambda k^{2}}; \quad a_{2}=\dfrac{p}{2\Lambda\tilde{s}^{1/4}};  \nonumber \\ 
a_{3}=\dfrac{q}{3\Lambda\tilde{s}^{1/2}}; \quad a_{4}=\dfrac{r}{4\Lambda\tilde{s}^{3/4}}; \quad \tilde{s}=\dfrac{s}{5\Lambda}, \label{arthur1}
\end{align}
with $\Lambda=k^{2}-\delta v^{2}$. 

The density-wave equation given in Eq. (\ref{ferrer2}) constitutes a generalized Liénard-type equation with higher-order nonlinearities, thereby providing a natural framework for the emergence of nontrivial travelling-wave solutions. It is worth noting that the reduced form of the improved HJ model, obtained by neglecting the $\epsilon$ and $\lambda$ terms in Eq. (\ref{ferrer1}), has been extensively studied in the literature over a wide range of nonlinear elastic parameters, including regimes lacking physical relevance. In the present work, however, we restrict our analysis to parameter ranges appropriate for biomembranes, namely $p<0$ and $q>0$ in Eq. (\ref{arthur1}). These values correspond to a nonlinear equation of state capable of supporting density-driven phase transitions between the fluid and gel phases, a feature encoded in the polynomial structure of Eq. (\ref{ferrer2}). A detailed discussion of the physical significance and values of these parameters can be found in \cite{tejocote38}.

In particular, the polynomial structure of Eq. (\ref{ferrer2}) admits a factorization procedure that underlies a supersymmetric pairing of solutions, among them Lambert $W$–kink solitons. Prior to constructing this supersymmetric framework, we examine the integrability properties of Eq. (\ref{ferrer2}) through the Painlevé test in the Kovalevskaya form, as originally proposed by S. V. Kovalevskaya \cite{tejocote045,tejocote046}.

\subsection{Painlev\'e test of integrability}
The Painlevé analysis begins by assuming a local solution of Eq.~(\ref{ferrer2}) in the form of a Laurent series 
\begin{equation}
y(\xi)=\sum_{\bar{k}=0}^{\infty}\bar{a}_{\bar{k}}(\xi-\xi_{0})^{\bar{k}-\bar{p}}, \label{elias1}
\end{equation}
where $\xi_{0}$ denotes the location of the movable singularity, $\bar{p}$ characterizes the leading-order behaviour, and $\bar{a}_{\bar{k}}$ are expansion coefficients. 

The Painlevé test can be summarized in the following three steps \cite{tejocote0046}:
\begin{enumerate}
\item We determine the leading-order behaviour of the solution near a movable singularity by performing a dominant balance \footnote{If the leading exponent $\bar{p}$ is not an integer, the expansion becomes of Puiseux type, indicating the presence of movable branch points rather than poles; consequently, the equation fails the Painlevé test.}.
\item We compute the Fuchs indices (resonances) associated with this leading-order behaviour.
\item We substitute the corresponding Laurent expansion into the original equation to verify the consistency of the recursion relations at the resonance levels.
\end{enumerate}

To implement this procedure, we first shift the singularity to the origin by introducing $\xi \to \xi-\xi_{0}$. We then seek the dominant behaviour in the form
\begin{equation}
y(\xi)=\bar{a}_{0}\xi^{-\bar{p}}, \label{annbre1}
\end{equation}
and substitute it into the dominant part of Eq. (\ref{ferrer2}), namely
\begin{equation}
\dfrac{d ^{2} {y}}{d\xi ^{2}}-y ^{5}=0. \label{annbre2}
\end{equation}
This yields $(\bar{a}_{0}, \bar{p})=\left(\pm \sqrt[4]{\dfrac{3}{4}}, \dfrac{1}{2}\right)$. Since $\bar{p}=1/2$ is non-integer, the movable singularity is a branch point rather than a pole. Consequently, Eq. (\ref{ferrer2}) fails the Painlevé test and is therefore non-integrable in the Painlevé sense. This suggests that globally meromorphic solutions are not expected in general and that alternative analytical techniques may be required.

Although the failure occurs already at the level of the leading-order analysis, it is still informative to compute the associated resonances, as they provide insight into the structure of the local (generally multivalued) solutions. The Painlevé test provides information about the analytic structure of solutions rather than a definitive criterion for solvability. 

In particular, the Fuchs indices (resonances) indicate the positions at which arbitrary constants may enter the local expansion and whether compatibility conditions are satisfied. If the leading-order exponent and all resonances are integer-valued, and the recursion relations are compatible at every resonance level, then the equation may possess the Painlevé property and admit locally single-valued expansions around movable singularities. 

To determine the associated resonances, we substitute the values $\bar{a}_{0}$ and $\bar{p}$ obtained from the preceding step in Eq. (\ref{elias1}) to yield 
\begin{equation}
y(\xi)=\pm \sqrt[4]{\dfrac{3}{4}}\xi^{-1/2}+\bar{a}_{j}\xi^{j-1/2}. \label{miranda1}
\end{equation}
Substituting Eq. (\ref{miranda1}) into Eq. (\ref{annbre2}) and collecting terms linear in $\bar{a}_j$ yields   
\begin{equation}
j^{2}-2j-3=0
\end{equation}
and, consequently, the resonances are $j_{1}=-1$ and $j_{2}=3$. The resonance $j=-1$ corresponds to the arbitrariness of the singularity position $\xi_{0}$, while $j=3$ indicates the presence of a free parameter entering the local Puiseux expansion. Despite the non-integer leading exponent, these resonances suggest a structured local behaviour of solutions near movable singularities. From this result, it is clear that the coefficient $\bar{a}_{3}$ remains arbitrary in the local Puiseux expansion. The failure of the Painlevé property indicates that Eq. (\ref{ferrer2}) is not integrable in the Painlevé sense. Nevertheless, this does not exclude the existence of particular exact solutions. Instead, it suggests that alternative approaches, such as suitable ansätze \cite{tejocote047}, parameter constraints, or factorization methods, are more appropriate for constructing explicit solutions. In this context, the emergence of Lambert $W$-function solutions is particularly natural, since the branch-point structure of the Lambert $W$ function mirrors the multivalued local behaviour predicted by the Painlevé analysis. 

\section{Electromechanical waves in the nerve membranes}
\subsection{The factorization method}
As previously mentioned, the factorization method provides a systematic and efficient framework for treating ordinary differential equations with polynomial nonlinearities that are not integrable in the Painlevé sense, such as Eq.~(\ref{ferrer2}). Its effectiveness has been demonstrated in a wide range of nonlinear problems whose governing equations possess a polynomial structure. In contrast to quasi-analytical techniques and approaches based on the inverse scattering transform, the method is more straightforward to implement while still capturing the essential features of the nonlinear dynamics. Consequently, it has found numerous applications in gravitation, biophysics, integrability theory, and mathematical physics \cite{tejocote45, tejocote46, tejocote47, tejocote48}.

To outline the method, consider a nonlinear differential equation of the form
\begin{equation}
\dfrac{d^{2}y}{d\xi^{2}} + \tilde{\gamma}\dfrac{dy}{d\xi} + f(y) = 0,
\label{ferrer3}
\end{equation}
which can be interpreted as a damped nonlinear oscillator, where $\tilde{\gamma}$ is a dissipation parameter and $f(y)$ is a nonlinear polynomial function.
The key idea consists in factorizing Eq.~(\ref{ferrer3}) as a product of first-order differential operators:
\begin{equation}
\left[\dfrac{d}{d\xi}-\phi_{2}(y)\right]\left[\dfrac{d}{d\xi}-\phi_{1}(y)\right]y=0.
\label{ferrer4}
\end{equation}
Expanding Eq.~(\ref{ferrer4}) and matching coefficients with Eq.~(\ref{ferrer3}) leads to the consistency conditions
\begin{subequations}
\begin{equation}
\phi_{1}(y)\phi_{2}(y)=\dfrac{f(y)}{y},
\label{ferrer5}
\end{equation}
\begin{equation}
\phi_{1}(y)+\phi_{2}(y)+y\dfrac{d\phi_{1}}{dy}=-\tilde{\gamma}.
\label{ferrer6}
\end{equation}
\end{subequations}
A particularly tractable class of solutions is obtained by imposing the first-order compatibility condition
\begin{equation}
\left[\dfrac{d}{d\xi}-\phi_{1}(y)\right]y=0,
\label{coyote1}
\end{equation}
which reduces the original second-order equation to a nonlinear first-order equation. Although this choice corresponds to the simplest factorization branch, it already yields nontrivial solution families.

It is worth emphasizing that Eq.~(\ref{coyote1}) is not unique: alternative factorizations may generate different solution branches, potentially involving functions of $\xi$ alone or extended dependencies (e.g., $\xi$ and $t$) \cite{tejocote51, tejocote52}. This framework has been previously explored in the context of the Liénard equation, particularly in cases where commutativity of the factorizing operators leads to significant simplifications and, in special instances, to isochronous systems \cite{tejocote50}. 
\subsection{Lambert $W$-Kink-type solitons}

Following the ideas of our previous work \cite{tejocote38}, and in order to obtain directly the Lambert $W$-Kink solitons, we consider the following form of Eq. (\ref{ferrer2}):
\begin{equation}
\dfrac{d^{2}y}{d\xi^{2}}+\tilde{\gamma}\dfrac{dy}{d\xi}+(y-\alpha) ^{2}(-y ^{2}+Ay+B)y=0, \label{draco1}
\end{equation}
where $A$, $B$ and $\alpha$ are obtained by solving the following overdetermined system of equations:
\begin{subequations}
\begin{equation}
-a_{4}=A+2\alpha; \label{better1}
\end{equation}
\begin{equation}
-a_{3}=B-2\alpha A-\alpha ^{2}; \label{better2}
\end{equation}
\begin{equation}
a_{2}=-2\alpha B +A\alpha ^{2}; \label{better3}
\end{equation}
\begin{equation}
-a_{1}=\alpha^{2}B. \label{better4}
\end{equation}
\end{subequations}
Moreover, since $\alpha$ appears explicitly in Eqs. (\ref{better1})-(\ref{better4}), the coefficients $A$ and $B$ can be parametrized in terms of this quantity, so that determining $\alpha$ completely characterizes the factorized polynomial structure. It is particularly convenient to express $\alpha$ solely in terms of the coefficients $a_{2}$, $a_{3}$ and $a_{4}$, as these parameters are directly associated with the elastic properties of the nerve membrane. To this end, substituting Eqs. (\ref{better1}) and (\ref{better2}) into Eq. (\ref{better3}) yields the following cubic algebraic equation for $\alpha$:
\begin{equation}
\alpha^{3}+\dfrac{3}{4}a_{4}\alpha^{2}+\dfrac{1}{2}a_{3}\alpha-\dfrac{1}{4}a_{2}=0, \label{better5}
\end{equation}  
and generally admits multiple roots. In the present work, we restrict our attention to real and non-vanishing values of $\alpha$ to exclude complex and trivial solutions that lack a direct physical interpretation. The nature of the roots of Eq. (\ref{better5}) is fully determined by its discriminant. To characterize the admissible parameter regions, we employ Cardano's method and introduce the invariants $P$ and $Q$, defined as
\begin{equation}
P=\dfrac{1}{2}a_{3}-\dfrac{3}{16}a_{4} ^{2}\label{hola1}
\end{equation}
and
\begin{equation}
Q=\dfrac{1}{32}a_{4}^{3}-\dfrac{1}{8}a_{4}a_{3}-\dfrac{1}{4}a_{2}\label{hola2}
\end{equation}
where $a_{i}$ ($i=1, 2, 3, 4$) can be expressed in terms of the original elastic coefficients by means of Eqs. $(\ref{arthur1})$. Additionally, we express its discriminant as:
\begin{equation}
D=\left(\dfrac{Q}{2}\right)^{2}+\left(\dfrac{P}{3}\right) ^{3},
\end{equation}
According to the trichotomy of the cubic discriminant, we know that $D$ may be either positive ($D>0$), negative ($D<0$) or zero ($D=0$), which yields one real root, three real roots or multiple real root (double or triple). The condition $D<0$ guarantees three distinct real solutions for the parameter $\alpha$. Through the factorization (\ref{draco1}), these solutions determine distinct real configurations of the polynomial nonlinearity and therefore generate multiple admissible equilibrium structures for the reduced dynamical system. Such multistability is a necessary condition for the construction of heteroclinic trajectories connecting different asymptotic states \cite{mafer00}. Since Lambert $W$-kink solitons arise precisely from these heteroclinic connections, the region $D<0$ defines a necessary existence domain for this class of solutions. In contrast, the regime $D>0$ yields a single real solution for $\alpha$, corresponding to a monostable polynomial structure in which heteroclinic connections between distinct equilibrium states cannot be constructed. Consequently, the condition $D=0$ represents the coalescence of real roots and may therefore be interpreted as a critical threshold separating the soliton-supporting ($D<0$) and monostable ($D>0$) regimes. 
Thus, once an admissible real root for $\alpha$, given by the Cardano formula
\begin{equation}
\alpha=-\dfrac{a_{4}}{4}+\sqrt[3]{-\dfrac{Q}{2}+\sqrt{D}}+\sqrt[3]{-\dfrac{Q}{2}-\sqrt{D}},  \label{cortegana}
\end{equation}
has been selected, the coefficients $A$ and $B$ are uniquely determined, thereby completely specifying the factorized representation in (\ref{draco1}). As will become evident in the following sections, an additional constraint on $A$ and $B$ arises from the commutation relations between the differential factors. For the particular form of Eq.~(\ref{draco1}), preserving a constant damping coefficient in Eq.~(\ref{ferrer6}) restricts the nonlinear polynomial function $f(y)/y$, factorized as $\phi_{1}(y)\phi_{2}(y)$ in Eq. ($\ref{ferrer5}$), to two admissible factorizations.

\subsubsection{Case I.}
Under the choice of $\phi_{1}(y)$ and $\phi_{2}(y)$
\begin{equation}
\phi_{1}(y)=\pm \dfrac{1}{\sqrt{3}}(y-\alpha) ^{2}; \qquad \phi_{2}(y)=\pm \sqrt{3}(-y ^{2}+Ay+B),\label{rel1}
\end{equation} 
the first allowed factorization is defined. Consequently, $A$, $B$ and $\tilde{\gamma}_{1}$ can be obtained from the factorization condition ($\ref{ferrer6}$), yielding
\begin{subequations}
\begin{equation}
A=\dfrac{4}{3}\alpha;  \qquad B=-a_{3}+\dfrac{11}{3}\alpha^{2}; \label{AB1}
\end{equation}
and
\begin{equation}
\tilde{\gamma}_{1}=\pm \sqrt{3}\left[a_{3}-4\alpha^{2}\right].  \label{gamma1}
\end{equation}
\end{subequations}
In addition, from the compatibility condition, given by Eq. (\ref{coyote1}), we have
\begin{equation}
\dfrac{dy}{d\xi}=\pm \dfrac{1}{\sqrt{3}}(y-\alpha)^{2}y  
\end{equation}
and, upon integrating, we obtain the following form of the Lambert $W$-kink soliton
\begin{equation}
y_{1}^{(1)}(\xi)=\alpha\left(1-\dfrac{1}{1+W\left[\varphi_{1}(\xi)\right]}\right), \label{Wkink1}
\end{equation}
where 
\begin{equation}
\varphi_{1} (\xi)= \exp \left(\pm \dfrac{\alpha ^{2}}{\sqrt{3}}\xi-1\right) \label{rodrigo1}
\end{equation}
with $W(\varphi_{1}(\xi))$ being the Lambert $W$ function, defined as the inverse function of $f(W)=We^{W}$ and $\alpha$ is determined by the real roots of Eq. (\ref{cortegana}), whose graphical representation is presented in Fig. \ref{jm1}. The expressions of Eq. (\ref{Wkink1}) for large values of $\xi$ are 
\[ \begin{cases} 
      y_{1}^{(1)}(\xi) \to \alpha \left(1 - \dfrac{\sqrt{3}}{\alpha^{2}\xi}\right) & \text{if} \quad  \xi \to \infty,  \\ 
      \\
     y_{1}^{(1)}(\xi)\to \alpha \exp\left(\dfrac{\alpha^{2}}{\sqrt{3}}\xi -1\right)  &  \text{if} \quad \xi \to -\infty, \\
   \end{cases}
\]
which are consistent with those obtained within the framework of the $\phi^{6}$ model. In the large-$\xi$ regime, the Lambert $W$-kink soliton exhibits a long-range, power-law decay, in contrast to the opposite side, where the asymptotic behaviour is exponential \cite{tejocotepre39}. It is clear that the asymptotic behaviour of the Lambert $W$-kink soliton in the limit of large $\xi$ is determined by the choice of sign in $\varphi_{1} (\xi)$, given by Eq. (\ref{rodrigo1}), which effectively interchanges the corresponding asymptotic expressions. 

\subsubsection{Case II.}
Alternatively, choosing the inverted order of \(\phi_{1}(y)\) and \(\phi_{2}(y)\) from Eq. (\ref{ferrer5}), with explicit forms obtained from the factorization of Eq. (\ref{draco1}), yields 
\begin{equation}
\phi_{1}(y)=\pm \dfrac{1}{\sqrt{3}}(-y ^{2}+Ay+B); \qquad \phi_{2}(y)=\pm \sqrt{3}(y-\alpha) ^{2}. \label{rel2}
\end{equation} 
Consequently, the factorization condition $(\ref{ferrer6})$ determines $A$, $B$ and $\tilde{\gamma}_{2}$ as
\begin{subequations}
\begin{equation}
A=3\alpha;  \qquad B=-a_{3}+7\alpha^{2} \label{AB2}
\end{equation}
and
\begin{equation}
\tilde{\gamma}_{2}=\pm \dfrac{1}{\sqrt{3}}\left[a_{3}-10\alpha^{2}\right].\label{gamma2}
\end{equation}
\end{subequations}
Thus, the compatibility condition, given by Eq. (\ref{coyote1}), can be expressed as
\begin{equation}
\dfrac{dy}{d\xi}=\pm \dfrac{1}{\sqrt{3}}(-y^{2}+Ay+B)y,
\end{equation}
which yields three distinct solution classes, depending on the parameter values.
\begin{subequations}
\begin{equation}
\dfrac{1}{B}\ln\left[ \dfrac{y_{2>} ^{(1)}}{\sqrt{-\left(y_{2>}^{(1)}\right)^{2}+Ay_{2>} ^{(1)}+B}}\right]-\dfrac{A}{\sqrt{\Delta}B}\text{arctanh}\left(\dfrac{2y_{2>} ^{(1)}-A}{\sqrt{\Delta}}\right)=\pm \dfrac{1}{\sqrt{3}}(\xi-\xi_{0}), \quad \text{if } \quad \Delta>0; \label{eqn1}
\end{equation}
\begin{equation}
\dfrac{1}{B}\ln\left[ \dfrac{y_{2<} ^{(1)}}{\sqrt{-\left(y_{2<} ^{(1)}\right)^{2}+Ay_{2<} ^{(1)}+B}}\right]+\dfrac{A}{\sqrt{\Delta}B}\arctan\left(\dfrac{2y_{2<} ^{(1)}-A}{\sqrt{\Delta}}\right)=\pm \dfrac{1}{\sqrt{3}}(\xi-\xi_{0}), \quad \text{if} \quad \Delta<0; \label{eqn2}
\end{equation}
\begin{equation}
 y_{2=} ^{(1)}(\xi)=\dfrac{3\alpha}{2}\left(1-\dfrac{1}{1+W\left[\varphi_{2}(\xi)\right]}\right) \quad \text{with} \quad  \varphi_{2}(\xi)=\exp \left(\mp\dfrac{ 3\sqrt{3}}{4}\alpha^{2}\xi-1\right), \quad \text{if } \quad \Delta=0; \label{eqn3}
\end{equation}
\end{subequations}
where $\Delta=A^{2}+4B$, denotes the discriminant of the quadratic polynomial $-y^{2}+Ay+B$, with $A$ and $B$ given by Eqs. (\ref{AB2}), which explicity can be expressed as $\Delta=\alpha ^{2}-{4a_{3}}/{37}$.  

As it is evident from the case yielded by Eq. (\ref{eqn3}) where $\Delta=0$, the arising of the Lambert $W$-kink-type solitons is not restricted to the particular choice of a given form of the polynomial nonlinearity in Eq. (\ref{draco1}), the restrictions to yield Lambert $W$-kink solitons will appear directly as a particular case of the product of two second order general polynomial of the form $A_{2}y^{2}+A_{1}y+A_{0}$ such as the mentioned in \cite{tejocote38} (footnote pag. 8). The other remaining cases $\Delta>0$ and $\Delta<0$, depicted in Figs. \ref{josemourinho1}a and \ref{josemourinho1}b,  for which we obtain the transcendental functions Eq. (\ref{eqn1}) and Eq. (\ref{eqn2}) are known solutions of the Abel equation of first type with constant coefficients. As it can be seen represented graphically these solutions will lead to a kink-like behaviour (Eq. (\ref{eqn1})) or to a periodic behaviour (\ref{eqn2}) for $\Delta >0$ and $\Delta<0$, correspondingly.   

Again for large values of $\xi$ the Lambert $W$-kink soliton, expressed by Eq. (\ref{eqn3}) and depicted in Fig. \ref{jm2}, is
\[ \begin{cases} 
      y_{2=} ^{(1)}(\xi) \to \dfrac{3}{2}\alpha \left(1 - \dfrac{4}{3\sqrt{3}\alpha^{2}\xi}\right) & \text{if} \quad  \xi \to \infty,  \\ 
      \\
     y_{2=} ^{(1)}(\xi) \to \dfrac{3}{2}\alpha \exp\left(\dfrac{3\sqrt{3}}{4}\alpha^{2}\xi -1\right)  &  \text{if} \quad \xi \to -\infty, \\
   \end{cases}
\]
which are similar to the previously obtained for the Lambert $W$-kink soliton of Case I. 

\begin{figure}[htbp]
    \centering
    \stackinset{r}{15pt}{t}{15pt}{\small {(b)}}{%
        \stackinset{l}{1 pt}{t}{10pt}{\small (a)}{%
            \stackinset{r}{10pt}{t}{10pt}{%
                \includegraphics[width=4.2cm]{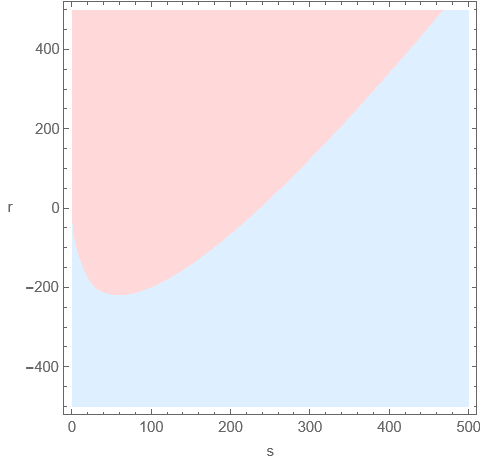}
            }{%
                \includegraphics[width=13cm]{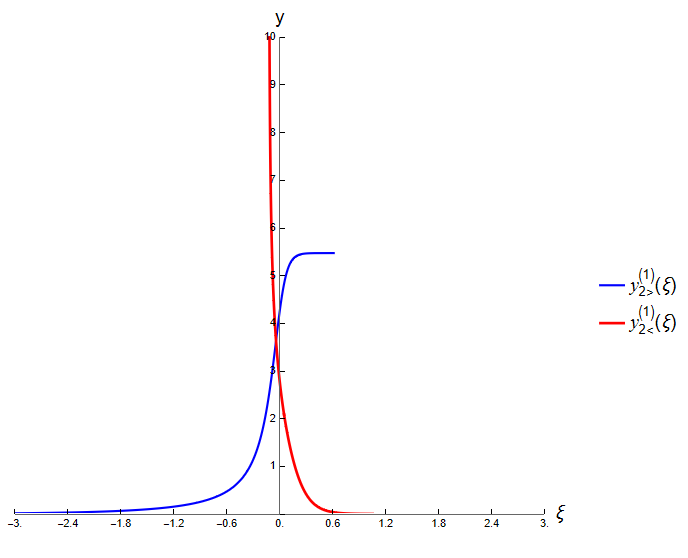}
            }%
        }%
    }
    \caption{(a) Graphical representation of the transcendental functions corresponding to $\Delta>0$ ($y_{2>} ^{(1)}(\xi)$) and $\Delta <0$ ($y_{2<} ^{(1)}(\xi)$) provided by Eqs. (\ref{eqn1}) and (\ref{eqn2}), correspondingly. The chosen parameters for $y_{2>} ^{(1)}(\xi)$ are $p=q=500$, $s=135$, $r=-250$, $k=2$, $\delta=v=1$ and $\xi_{0}=0$ and for $y_{2<} ^{(1)}(\xi)$ are $p=q=500$, $s=60$, $r=-170$, $k=2$, $\delta=v=1$ and $\xi_{0}=0$. (b) Region of existence at $(r, s)$ for $y_{2>} ^{(1)}(\xi)$ and $y_{2<} ^{(1)}(\xi)$. } \label{josemourinho1}
\end{figure} 
We remark that, throughout this and the following section, we adopt the simplified notation $y ^{(1)}_{1>}$. In accordance with the standard supersymmetric quantum mechanics (SUSY QM) convention, the first subscript labels the solution corresponding to case I or II, the second subscript indicates the relevant branch when applicable, and the superscript specifies the associated potential. In the present section, the superscript (1) refers to the potential defined by Eq. (\ref{draco1}); the supersymmetric partner solutions associated with the second potential will be denoted by the superscript (2). 
\subsection{Supersymmetric pairing of Lambert $W$-kink-type solitons}
In the previous section, we have implicitly shown that the factorization conditions Eqs. $(\ref{ferrer5})$ and $(\ref{ferrer6})$ are not commutative. Thus, the ordering of the differential operators in Eq. (\ref{ferrer4}) is essential. Consequently, a direct reversing of $\phi_{1}(y)$ and $\phi_{2}(y)$, maintaining the damping coefficient constant, either $\tilde{\gamma}_{1}$ for the case I or $\tilde{\gamma}_{2}$ for the case II, will lead to a system described by a completely different equation. This property was termed, by the authors of the factorization method for nonlinear differential equations, as \textit{supersymmetric pairing} \cite{tejocote41}. However, their analysis was restricted to conventional kink solitons, without exploring in detail the physical interpretation of the associated partner equations. Therefore, in the present section we extend this idea to the Lambert $W$-kink solitons obtained from the extended HJ model and offer a possible physical interpretation relevant to traumatic brain injury for the additional terms that appear.  
\subsubsection{Case I.}
By direct reversing $\phi_{1}(y)$ and $\phi_{2}(y)$, given by Eqs. (\ref{rel1}), namely, 
\begin{equation}
\phi_{1}(y)=\pm \sqrt{3}(-y ^{2}+Ay+B); \qquad \phi_{2}(y)=\pm \dfrac{1}{\sqrt{3}}(y-\alpha) ^{2}, \label{newo1}
\end{equation} 
and $A$, $B$ and $\tilde{\gamma}_{1}$,  given by Eqs. (\ref{AB1}) and Eq. (\ref{gamma1}). The compatibility condition (\ref{coyote1}) now reads as: 
\begin{equation}
\dfrac{dy}{d\xi}=\pm \sqrt{3}(-y ^{2}+Ay+B)y.
\end{equation}
Thus, depending on the values of the constants \(A\) and \(B\) defined in (\ref{AB1}), we obtain three solutions:
\begin{subequations}
\begin{equation}
\dfrac{1}{B}\ln\left[ \dfrac{y_{1>} ^{(2)}}{\sqrt{-\left(y_{1>} ^{(2)}\right) ^{2}+Ay_{1>} ^{(2)}+B}}\right]-\dfrac{A}{\sqrt{\Delta}B}\text{arctanh}\left(\dfrac{2y_{1>} ^{(2)}-A}{\sqrt{\Delta}}\right)=\pm \sqrt{3}(\xi-\xi_{0}), \quad  \text{if } \quad \Delta>0; \label{eqn1a}
\end{equation}
\begin{equation}
\dfrac{1}{B}\ln\left[ \dfrac{y_{1>} ^{(2)}}{\sqrt{-\left(y_{1>} ^{(2)}\right) ^{2}+Ay_{1>} ^{(2)}+B}}\right]+\dfrac{A}{\sqrt{\Delta}B}\arctan\left(\dfrac{2y_{1>} ^{(2)}-A}{\sqrt{\Delta}}\right)=\pm \sqrt{3}(\xi-\xi_{0}), \quad  \text{if } \quad \Delta<0; \label{eqn2a}
\end{equation}
\begin{equation}
y_{1=} ^{(2)}(\xi)=\dfrac{2\alpha}{3}\left(1-\dfrac{1}{1+W\left[\varphi_{1'}(\xi)\right]}\right) \quad \text{with} \quad  \varphi_{1'}(\xi)=\exp \left(\mp\dfrac{ 4}{3\sqrt{3}}\alpha^{2}\xi-1\right), \quad \text{if } \quad \Delta=0 \label{eqn3a}
\end{equation}
\end{subequations}
with $\Delta=A^{2}+4B$, where $A$ and $B$ are defined in Eqs.~(\ref{AB1}). The solutions given by Eqs.~(\ref{eqn1a}), (\ref{eqn2a}) and (\ref{eqn3a}) constitute supersymmetric partners of the Lambert $W$-kink soliton (\ref{Wkink1}).

To make the supersymmetric pairing more transparent in its original formulation, namely, by considering the same wavefront velocity in the diffusion equation (\ref{ferrer3}), we focus on the Lambert $W$-kink solutions given by Eqs.~(\ref{Wkink1}) and (\ref{eqn3a}). This representation enables a direct comparison of how the fundamental soliton parameters are transformed between the two supersymmetric partners. As shown in Fig.~\ref{jm1}, both Lambert $W$-kink solutions preserve the damping coefficient $\tilde{\gamma}$, which is associated with the axoplasmic fluid, while differing only in the direction of propagation.

To compare solitons propagating in the same direction, opposite sign conventions must be adopted in Eqs.~(\ref{Wkink1}) and (\ref{eqn3a}). The resulting correspondence is illustrated schematically in Fig.~\ref{charmander}. Under this convention, the supersymmetric transformation produces significant changes in the soliton amplitude and width. By contrast, the supersymmetric partner solutions of the Lambert $W$-kink soliton (\ref{Wkink1}), given by Eq. (\ref{eqn1a}) for $\Delta >0$ and Eq.~(\ref{eqn2a}) for $\Delta <0$, exhibit a considerably stronger deformation. As illustrated in Figs.~\ref{josemourinho2}a and \ref{josemourinho2}b, this behaviour reflects the deformation of the corresponding effective potential and is manifested through their transcendental profiles and regions of existence.
\begin{figure}[htbp]
    \centering
    \stackinset{r}{15pt}{t}{15pt}{\small {(b)}}{%
        \stackinset{l}{1 pt}{t}{10pt}{\small (a)}{%
            \stackinset{r}{10pt}{t}{10pt}{%
                \includegraphics[width=4.2cm]{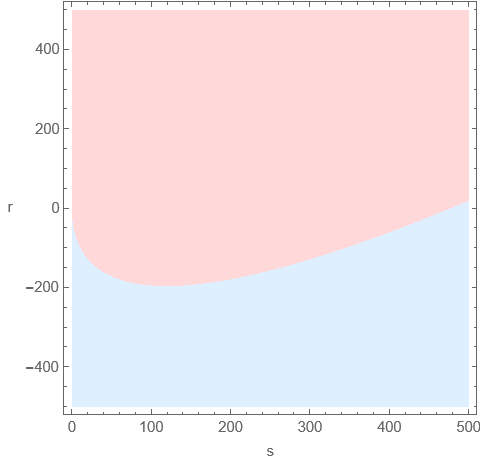}
            }{%
                \includegraphics[width=13cm]{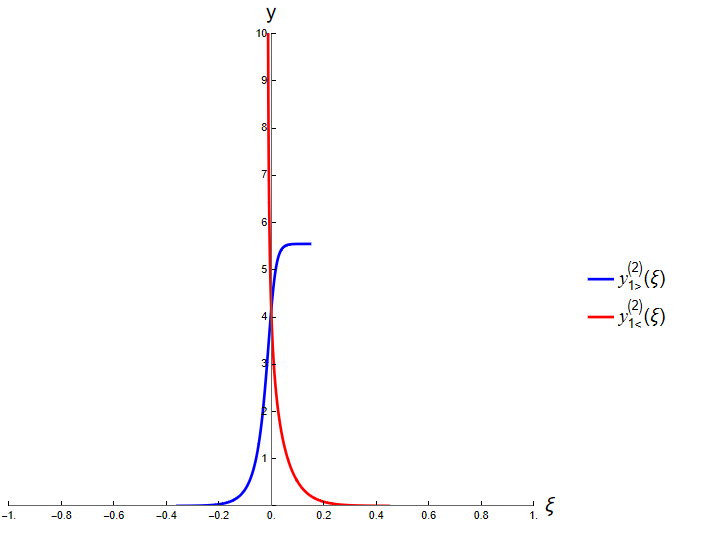}
            }%
        }%
    }
    \caption{(a) Graphical representation of the transcendental functions corresponding to $\Delta>0$ ($y_{1>} ^{(2)}(\xi)$) and $\Delta <0$ ($y_{1<} ^{(2)}(\xi)$) provided by Eqs. (\ref{eqn1a}) and (\ref{eqn2a}), correspondingly. The chosen parameters for $y_{1>} ^{(2)}(\xi)$ are $p=q=300$, $s=250$, $r=-400$, $k=2$, $\delta=v=1$ and $\xi_{0}=0$ and for $y_{1<} ^{(2)}$ are $p=q=300$, $s=100$, $r=200$, $k=2$, $\delta=v=1$ and $\xi_{0}=0$. (b) Region of existence at $(r, s)$ for $y_{1>} ^{(2)}(\xi)$ and $y_{1<} ^{(2)}(\xi)$.} \label{josemourinho2}
\end{figure} 

This modified effective potential can be obtained through a direct substitution of Eq.~(\ref{newo1}) into Eq.~(\ref{ferrer4}), with $\tilde{\gamma}_{1}$ given by Eq.~(\ref{gamma1}). The resulting expression yields the supersymmetric partner equation, which can be written as
\begin{equation}
\dfrac{d^{2}y}{d \xi^{2}} \mp \tilde{\gamma}_{1} \dfrac{dy}{d\xi}+(9y^{2}-8\alpha y+\alpha ^{2})(-y^{2}+Ay+B)y=0. \label{rob1}
\end{equation}
In general, this reversing in the order of $\phi_{1}(y)$ and $\phi_{2}(y)$ and its corresponding partner equation, can be seen as a modulation of the original potential, which causes the potential to modified or, in a more drastic case, may be interpreted as an effective modulation of the membrane potential induced by external mechanical perturbations. Considering Eq. (\ref{rob1}), together with the mechanical analogy, we can find the potential $V_{1}(y)$ and the supersymmetric partner potential $V_{2}(y)$ for $\tilde{\gamma}_{1}$, given by Eq. (\ref{gamma1}), to have the explicit forms
\begin{subequations}
\begin{equation}
V_{1}(y)=\dfrac{1}{6}y^{6}-\dfrac{A+2\alpha}{5}y ^{5}+\dfrac{\alpha ^{2}+2A \alpha-B}{4}y ^{4}+\dfrac{2\alpha B-A\alpha^{2}}{3}y^{3} -\dfrac{B\alpha^{2}}{2}y^{2} \label{aa1}
\end{equation}
\begin{equation}
V_{2}(y)=\dfrac{3}{2}y^{6}-\dfrac{9A+8\alpha}{5}y ^{5}+\dfrac{\alpha ^{2}+8A \alpha-9B}{4}y ^{4}+\dfrac{8\alpha B-A\alpha^{2}}{3}y^{3} -\dfrac{B\alpha^{2}}{2}y^{2} \label{aa2}
\end{equation}
\end{subequations}
with $A$ and $B$ given by Eqs. (\ref{AB1}), correspondingly. Once we depict the potentials $V_{1}(y)$ and $V_{2}(y)$, see Fig. \ref{elemir1}, and considering the local minimum as phase transition. It is natural to interpret the first local minimum (located in the negative $y$-region) as the first phase, the second local minimum (located in the positive $y$-region) as the second phase, and the intermediate region as the transition front \cite{phtrans}. From this perspective, the commutation-induced reversal of the Lambert $W$-kinks, described by Eqs.~(\ref{Wkink1}) and (\ref{eqn3a}), corresponds to the reversal of the phase-transition front. Thus, the original gel-to-liquid transition remains intact, while its propagation direction is reversed. 

\begin{figure}[H]
\includegraphics[width=0.7\textwidth]{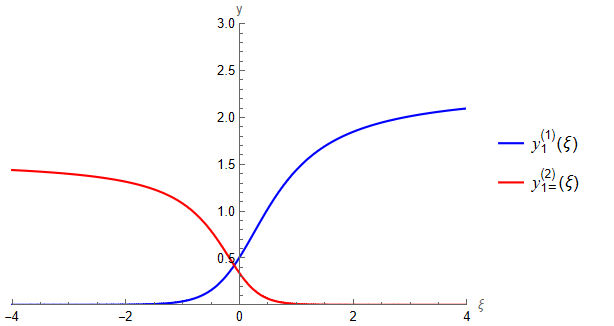}
\centering
\caption{Lambert $W$-kink soliton ($y_{1} ^{(1)}(\xi)$) and supersymmetric paired Lambert $W$-kink soliton ($y_{1=} ^{(2)}(\xi)$) given by Eq. (\ref{Wkink1}) and Eq. (\ref{eqn3a}), correspondingly, with the chosen parameters $p=q=300$, $s=27$, $r=-141$, $k=2$, $\delta=v=1$.}\label{jm1} 
\end{figure}

\begin{figure}[H]
\centering
    \begin{subfigure}[b]{0.45\textwidth}
    \centering
    \includegraphics[width=8.7cm]{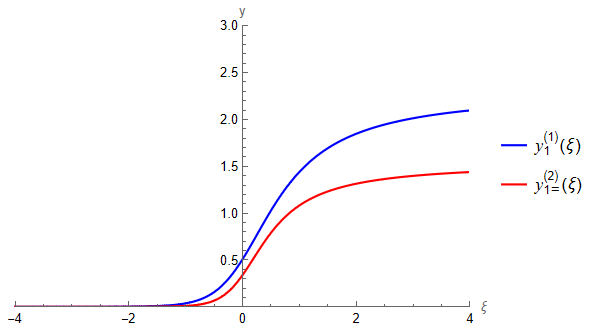}
    \caption{}
    \end{subfigure}
\quad
    \begin{subfigure}[b]{0.45\textwidth}
    \centering
    \includegraphics[width=9.3cm]{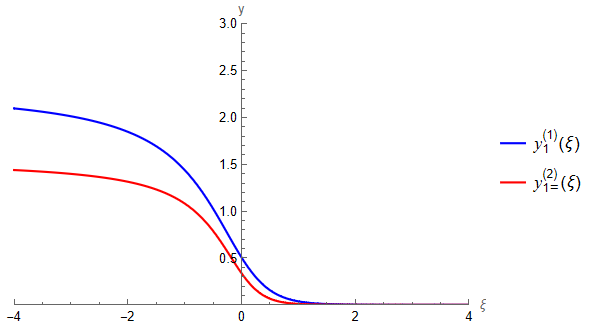}
    \caption{}
    \end{subfigure}
\caption{(a) Lambert $W$-kink soliton ($y_{1} ^{(1)}(\xi)$) and supersymmetric paired Lambert $W$-kink soliton ($y_{1=} ^{(2)}(\xi)$) given by Eq. (\ref{Wkink1}) and Eq. (\ref{eqn3a}) with interchanged signs, respectively, (b) Lambert $W$-kink soliton and supersymmetric paired Lambert $W$-kink soliton given by  Eq. (\ref{Wkink1}) and Eq. (\ref{eqn3a}) with interchanged signs with opposite direction, correspondingly. For (a) and (b) the chosen parameters are $p=q=300$, $s=27$, $r=-141$, $k=2$, $\delta=v=1$.} \label{charmander}
\end{figure}

\begin{figure}[H]
\includegraphics[width=0.9\textwidth]{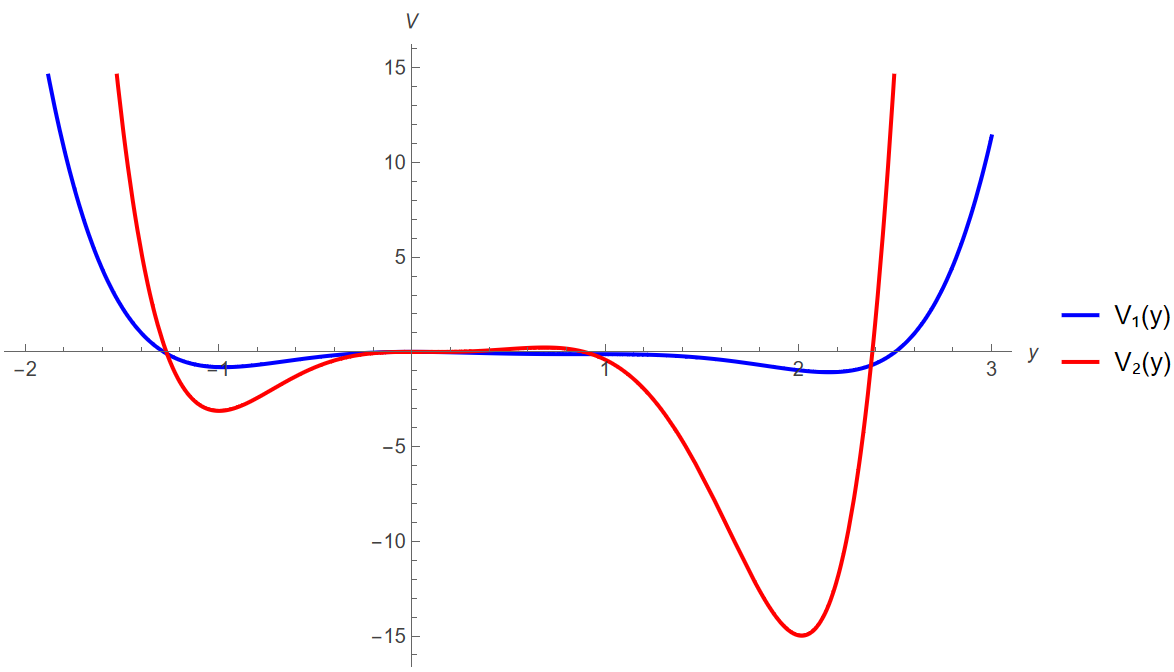}
\centering
\caption{Potential $V_{1}(y)$ and supersymmetric partner potential $V_{2}(y)$ given by Eq. (\ref{draco1}) and Eq. (\ref{rob1}), correspondingly. For both potentials the illustrative physical parameters were  $p=8$, $k=2$, $v=\delta=1, q=6$, $r=-9.5$ and $s=10.8$.} \label{elemir1} 
\end{figure}
It is worth emphasizing that the sextic potentials $V_{i}(y)$ with $i=1, 2$, defined by Eqs. (\ref{aa1}) and (\ref{aa2}), arise not only in the framework of the $\phi^{6}$ model but also in a variety of quantum-mechanical problems, including tunnelling, Wigner entropy, and the analysis of supersymmetric quantum states \cite{sexticpotential1, sexticpotential2, sexticpotential3}. Even these potentials seem to graphically satisfy the \textit{shape invariance} condition (see Fig. \ref{elemir1}), from the SUSY QM, defined as 
\begin{equation}
V_{2}(y; a_{1})=V_{1}(y; a_{2})+R(a_{1}), \label{invariant}
\end{equation}
where $a_{1}$ is a set of parameters, and $a_{2}=f(a_{1})$, and $R(a_{1})$ is independent of the variable $y$ \cite{susy1}, a direct computation shows that the potentials $V_{i}(y)$ are not shape invariant in a strict sense. This condition is essential to decide if the potentials do or do not belong to the same family of potentials. 
\subsubsection{Case II.}
Now, similar to previously analysed case, reversing the order of $\phi_{1}(y)$ and $\phi_{2}(y)$ in Eq. (\ref{rel2}) yields
\begin{equation}
\phi_{1}(y)=\pm \sqrt{3}(y-\alpha) ^{2}; \qquad \phi_{2}(y)=\pm \dfrac{1}{\sqrt{3}}(-y ^{2}+Ay+B). \label{newo2}
\end{equation} 
Consequently, $\tilde{\gamma}_{2}$, $A$ and $B$ are determined from the factorization conditions in Eq. (\ref{AB2}), 
\begin{equation}
\dfrac{dy}{d\xi}=\pm \sqrt{3}(y-\alpha) ^{2}y.
\end{equation}
Once we solve for $y$, we obtain
\begin{equation}
y_{2}^{(2)}(\xi)=\alpha \left(1-\dfrac{1}{1+W[\varphi_{2'}(\xi)]}\right), \label{wkink2}
\end{equation} 
where 
\begin{equation}
\varphi_{2'}(\xi)=\exp\left(\pm \sqrt{3}\alpha^{2}\xi-1\right),
\end{equation} 
and $W[\varphi_{2'}]$ is the Lambert $W$ function. 

The Lambert \(W\)-kink soliton (\ref{wkink2}) acts as the supersymmetric partner of (\ref{eqn3}), with modified physical properties such as width and amplitude. 	Clearly, for the supersymmetric pairs with $\Delta>0$ and $\Delta<0$, corresponding to Eq. (\ref{eqn2}) and Eq. (\ref{eqn3}), the Lambert $W$-kink soliton profile undergoes a complete modification. Similar to the previously analyzed case, plotting the Lambert \(W\)-kink soliton given by Eq.~(\ref{eqn3}) alongside its corresponding supersymmetric partner (\ref{wkink2}) reveals that both solutions preserve the same damping coefficient while their directions of propagation are reversed (see Fig.~\ref{jm2}). However, it is instructive to note that interchanging the sign conventions recovers both the Lambert \(W\)-kink soliton and its supersymmetric partner propagating in the same direction, as illustrated in Figs.~\ref{charmeleon}a and \ref{charmeleon}b. In this representation, the modifications to the soliton width and amplitude become clearly evident. 
\begin{figure}[H]
\includegraphics[width=0.7\textwidth]{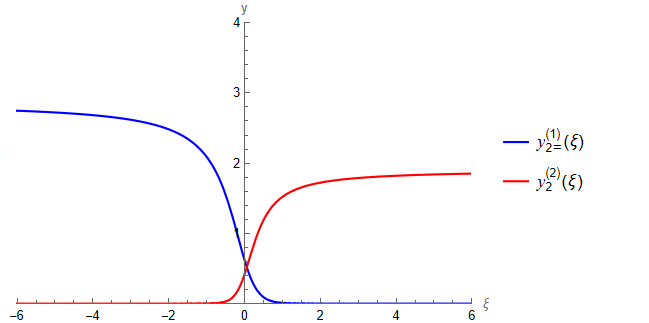}
\centering
\caption{Lambert $W$-kink soliton ($y_{2=}^{(1)}(\xi)$) and supersymmetric paired Lambert $W$-kink soliton ($y_{2}^{(2)}(\xi)$) given by Eq. (\ref{eqn3}) and Eq. (\ref{wkink2}), correspondingly, with the chosen parameters $p=q=500$, $s=437$, $r=-429$, $k=2$, $\delta=v=1$.}\label{jm2} 
\end{figure}
\begin{figure}[H]
\centering
    \begin{subfigure}[b]{0.45\textwidth}
    \centering
    \includegraphics[width=8.7cm]{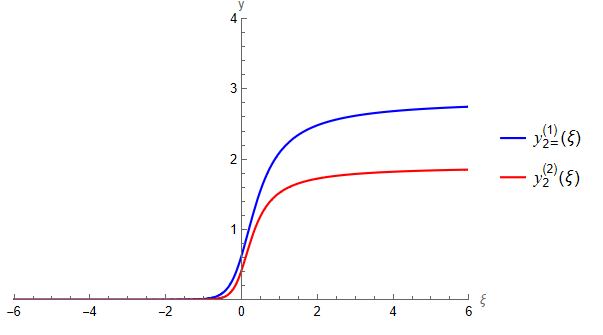}
    \caption{}
    \end{subfigure}
\quad
    \begin{subfigure}[b]{0.45\textwidth}
    \centering
    \includegraphics[width=9.3cm]{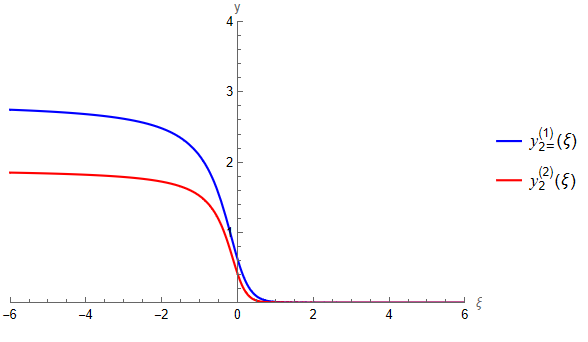}
    \caption{}
    \end{subfigure}
\caption{(a) Lambert $W$-kink soliton ($y_{2=}^{(1)}(\xi)$) and supersymmetric paired Lambert $W$-kink soliton ($y_{2}^{2}(\xi)$) given by Eq. (\ref{eqn3}) and Eq. (\ref{wkink2}) with interchanged signs, respectively, (b) Lambert $W$-kink soliton ($y_{2=}^{(1)}(\xi)$) and supersymmetric paired Lambert $W$-kink soliton ($y_{2}^{2}(\xi)$) given by Eq. (\ref{eqn3}) and Eq. (\ref{wkink2}) with interchanged signs with opposite direction, correspondingly. For (a) and (b) the chosen parameters are $p=q=500$, $s=437$, $r=-429$, $k=2$, $\delta=v=1$.} \label{charmeleon}
\end{figure}
Similar to the previous case, the direct inversion of $\phi_{1}(y)$ and $\phi_{2}(y)$, given by $(\ref{newo2})$, for the damping coefficient $\tilde{\gamma}_{2}$ leads to a Li\'enard equation with modulated potential:
\begin{equation}
\dfrac{d^{2}y}{d \xi^{2}} \mp \tilde{\gamma}_{2} \dfrac{dy}{d\xi}+(-9y^{2}+12\alpha y+B)(y-\alpha)^{2}y=0. \label{rob2}
\end{equation}
Again by mechanical analogy, we can determine the potential $V_{1}(y)$ alongside its supersymmetric partner potential $V_{2}(y)$ for the damping coefficient $\tilde{\gamma}_{2}$:
\begin{subequations}
\begin{equation}
V_{1}(y)=\dfrac{1}{6}y^{6}-\dfrac{A+2\alpha}{5}y ^{5}+\dfrac{\alpha ^{2}+2A \alpha-B}{4}y ^{4}+\dfrac{2\alpha B-A\alpha^{2}}{3}y^{3} -\dfrac{B\alpha^{2}}{2}y^{2} \label{aaa1}
\end{equation}
\begin{equation}
V_{2}(y)=\dfrac{3}{2}y^{6}-\dfrac{21\alpha}{5}y ^{5}+\dfrac{33\alpha^{2}-B}{4}y ^{4}+\dfrac{2\alpha B-12\alpha^{2}}{3}y^{3} -\dfrac{B\alpha^{2}}{2}y^{2} \label{aaa2}
\end{equation}
\end{subequations}
with $A$ and $B$ defined from Eqs. (\ref{AB2}). Similar to the case of potentials $V_{i}(y)$, with $i=1, 2$, of the previous subsection, we can see that the potentials $(\ref{aaa1})$ and $(\ref{aaa2})$ do not satisfy the shape invariance condition (\ref{invariant}).  
\begin{figure}[H]
\centering
    \begin{subfigure}[b]{0.45\textwidth}
    \centering
    \includegraphics[width=8.9cm]{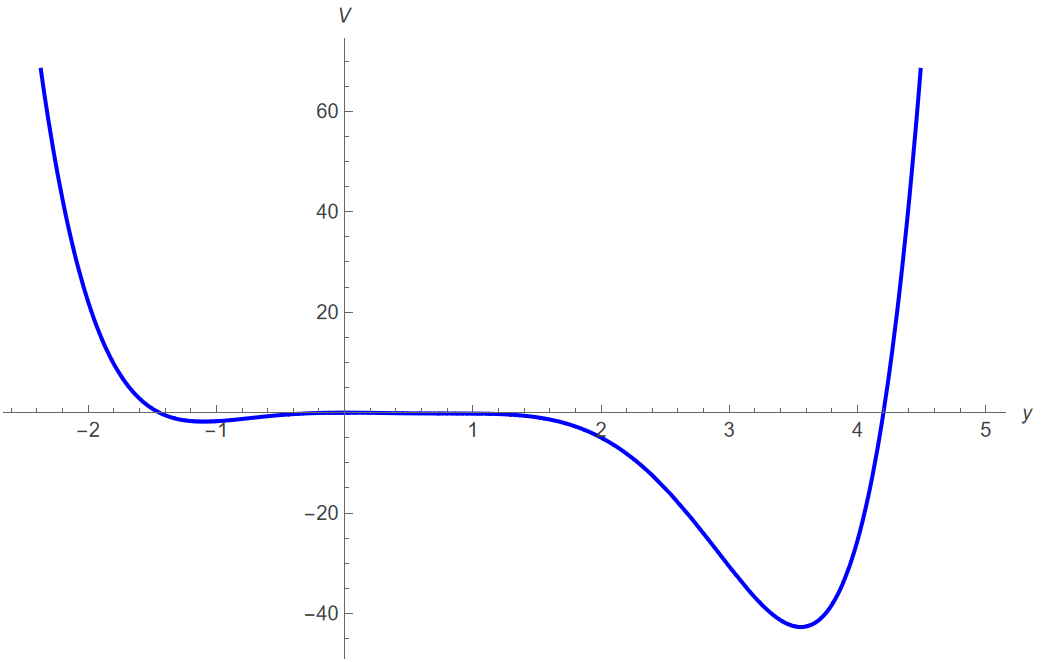}
    \caption{}
    \end{subfigure}
\quad
    \begin{subfigure}[b]{0.45\textwidth}
    \centering
    \includegraphics[width=8.9cm]{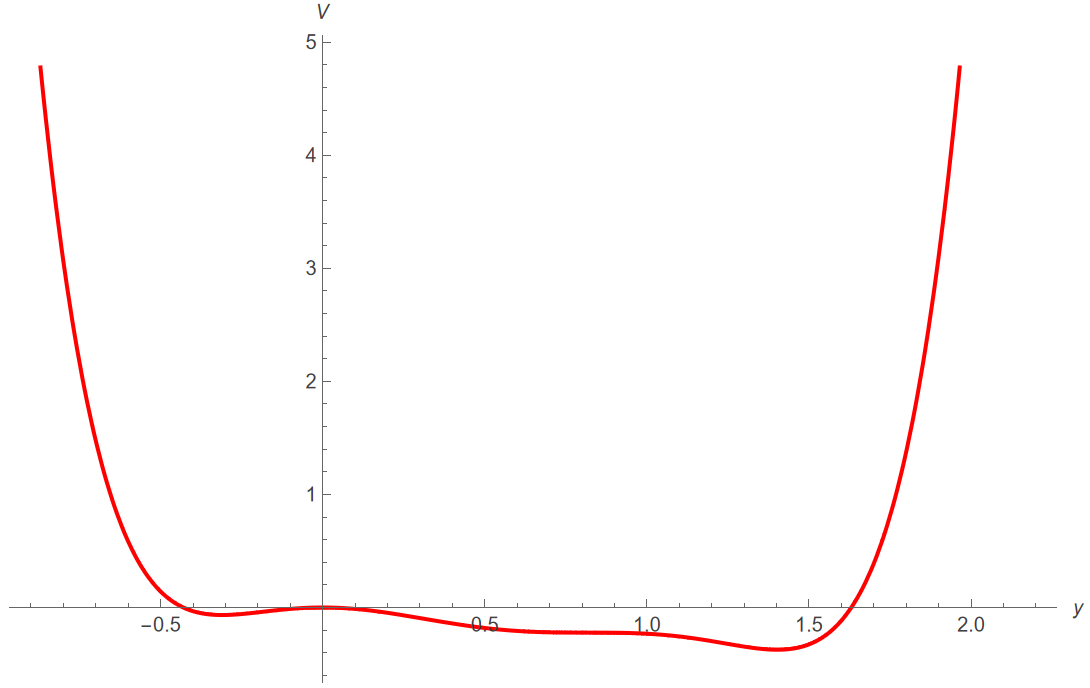}
    \caption{}
    \end{subfigure}
\caption{(a) Potential $V_{1}(y)$ given by Eq. (\ref{draco1}),  and (b) and supersymmetric partner potential $V_{2}(y)$ given by Eq. (\ref{rob2}). For both potentials the illustrative physical parameters were  $p=8$, $k=2$, $v=\delta=1, q=6$, $r=-9.5$ and $s=10.8$. } \label{charizard}
\end{figure}
We could delve deeper in the physiological significance of the supersymmetric paired potentials, given in Eqs. (\ref{aa2}) and (\ref{aaa2}), and its associated exact travelling wave solutions, interpreting them as alterations in the physiological conditions governing signal transmission, potentially associated with neural anomalies, injuries, or pathological damage \cite{tejocote43, tejocote44}. For instance, during seizure episodes or neuronal paroxysmal discharges, variations in pulse width and amplitude, together with partial distortion or breakdown of the propagating profiles, naturally arise. Such modifications of pulse width, amplitude, and profile shape are qualitatively reminiscent of alterations observed in abnormal neural activity. Although the present model does not directly describe specific neurological disorders, it suggests that changes in the effective nonlinear potential may provide a useful framework for exploring how pathological conditions influence electromechanical signal propagation \cite{feriha1, feriha2}.  
 
In particular, challenges remain in accurately accounting for myelination effects, membrane heterogeneity, and the mechanical properties of realistic biomembranes \cite{tejocote013, tejocote0013, tejocote00013}. These limitations, however, do not undermine the underlying physical picture; rather, they highlight the need for systematic extensions of the model. In this context, incorporating fractional-order dynamics together with position- and time-dependent coefficients provides a natural pathway toward a more realistic, non-autonomous generalization capable of capturing multiscale behaviour and memory effects inherent to neuronal systems. Such developments offer a promising direction for future research. 
As a final remark, it is important to emphasize that, independently of the validity of the extended HJ model analyzed in the present work, action potentials triggered by mechanosensory processes are not a new concept. Similar phenomena have long been observed in several biological systems, particularly in \textit{Mimosa pudica}, one of the best-known examples \cite{llamaeterna0,llamaeterna1}. Investigating these comparatively simpler systems could provide valuable physical insight into the mechanisms underlying nerve impulse generation and propagation in more complex excitable media.  
\section{Conclusion}
In this work, we considered an improved HJ model with strong polynomial nonlinearities, leading to a richer description of electromechanical wave dynamics in axonal membranes. In particular, the proposed extension naturally captures asymmetric pulse profiles, a feature that is not reproduced by the standard improved HJ model. This increased physical realism is accompanied by a significant mathematical consequence: the resulting Liénard equation loses its integrability in the Painlevé sense, thereby placing the analysis within a genuinely non-integrable regime.

Despite this loss of integrability, we demonstrated that exact travelling-wave solutions can still be obtained through the factorization method. The resulting Lambert $W$-kink solitons constitute a new class of analytical solutions for the extended HJ framework and reveal the persistence of coherent nonlinear structures under strong polynomial nonlinearities. Two distinct families of solutions were identified, each associated with different values of the axoplasmic fluid constant (or damping coefficient), thereby providing alternative dynamical regimes for pulse propagation.

Furthermore, by exploiting the analogy between factorization methods and SUSY QM, we constructed supersymmetric partner Liénard equations and their corresponding solutions. This approach generates modulated effective potentials, which are not shape invariant, and establishes a systematic mechanism for producing new waveforms from known solutions while preserving the underlying mathematical structure. The partner solutions exhibit significant variations in amplitude, width, and overall profile, while maintaining the same damping coefficient as the original waves. In extreme cases, the supersymmetric transformation leads to qualitatively distinct pulse morphologies, suggesting the existence of a broader family of electromechanical excitations than previously considered within the HJ framework.

The analytical and numerical investigation of the existence regions of these solutions further supports their robustness and reveals a rich parameter landscape. Moreover, a possible biological interpretation was proposed in which the supersymmetric potentials are associated with external mechanical loads or environmental perturbations acting on the membrane, while the partner solutions represent the corresponding cellular response. Within this perspective, the supersymmetric construction provides a mathematically consistent framework for studying how mechanical modulation may alter the characteristics of propagating nerve pulses.

More broadly, the present results demonstrate that Lambert $W$-kink solitons can emerge in non-integrable extensions of the HJ model and that supersymmetric techniques offer a powerful tool for generating and classifying families of electromechanical waves in nonlinear biological media. These findings open new avenues for investigating mechanically modulated nerve signals and their interactions with heterogeneous cellular environments. Future investigations should focus on the stability and experimental relevance of these supersymmetric electromechanical structures, as well as on their potential role in the regulation and adaptation of biological signal propagation in non-homogeneous media and in the presence of non-constant elastic coefficients.
\end{justify}

\section*{Acknowledgments}
OPT acknowledges SECIHTI for a postdoctoral fellowship. 

\medskip

\noindent \textbf{Data Availability Statement} Data sharing not applicable to this article as no datasets were generated or analyzed during the current study.


\begin{thebibliography}{9}
\bibitem{tejocote1} Drukarch B, Holland H A, Velichkov M, Geurts J J G, Voorn P, Glas G and de Regt H W 2018 Thinking about the nerve impulse: a critical analysis of the electricity-centered conception of nerve excitability Progress in Neurobiology 169 172185. 
\bibitem{tejocote2} Fields R D 2011 Signaling by neuronal swelling Science Signaling 4.
\bibitem{tejocote3} Tamm, K., Peets, T. \& Engelbrecht, J. The modelling of the action potentials in myelinated nerve fibres. Biomech Model Mechanobiol 25, 13 (2026).
\bibitem{tejocote4} Karami, G., Grundman, N., Abolfathi, N., Naik, A., Ziejewski, M., 2009. A micromechanical hyperelastic modeling of brain white matter under large deformation. J. Mech. Behav. Biomed. 2, 243–254.
\bibitem{tejocote5} Cloots, R.J.H., Nyberg, T., Kleiven, S., van Dommelen, J.A.W., Geers, M.G.D., 2011. Micromechanics of diffuse axonal injury: influence of axonal orientation and anisotropy. Biomech. Model. Mechanobiol. 10, 413–422.
\bibitem{tejocote6} Abolfathi, N., Naik, A., Chafi, M.S., Karami, G., Ziejewski, M., 2009. A micromechanical procedure for modelling the anisotropic mechanical properties of brain white matter. Comput. Method Biomec. 12, 249–262.
\bibitem{tejocote7} Faller R 2020 UCD Biophysics 241: Membrane Biology (LibreTexts).
\bibitem{tejocote07} Doman, E.A., Ovenden, N.C., Phillips, J.B. et al. Biomechanical modelling infers that collagen content within peripheral nerves is a greater indicator of axial Young’s modulus than structure. Biomech Model Mechanobiol 24, 297–309 (2025).
\bibitem{tejocote007} Smruta Koppaka, Allison Hess-Dunning, Dustin J. Tyler. Biomechanical characterization of isolated epineurial and perineurial membranes of rabbit sciatic nerve. Journal of Biomechanics 136 (2022) 111058. 
\bibitem{tejocote0007} Singh A, Kozin S and Balasubramanian S (2025) Biomechanical responses of peripheral nerves in human, pig and rat: a comparative study. Front. Bioeng. Biotechnol. 13:1641386. doi: 10.3389/fbioe.2025.1641386   
\bibitem{tejocote8} Hodgkin A L and Huxley A F 1952 A quantitative description of membrane current and its application to conduction and excitation in nerve The Journal of Physiology 117 500544. 
\bibitem{tejocote9} Hodgkin A L and Huxley A F 1952 Currents carried by sodium and potassium ions through the membrane of the giant axon of loligo The Journal of Physiology 116 449472.
\bibitem{tejocote10} FitzHugh R 1961 Impulses and physiological states in theoretical models of nerve membrane Biophysical Journal 1 445466. 
\bibitem{tejocote11} Nagumo J, Arimoto S and Yoshizawa S 1962 An active pulse transmission line simulating nerve axon Proceedings of the IRE 50 20612070. 
\bibitem{tejocote12} Heimburg T and Jackson A D 2007 On the action potential as a propagating density pulse and the role of anesthetics Biophysical Reviews and Letters 02 5778.
\bibitem{tejocote13} Andersen S S L, Jackson A D and Heimburg T 2009 Towards a thermodynamic theory of nerve pulse propagation Progress in Neurobiology 88 104113. 
\bibitem{tejocote14} El Hady A and Machta B B 2015 Mechanical surface waves accompany action potential propagation Nature Communications 6. 
\bibitem{tejocote15} Rvachev M M 2010 On axoplasmic pressure waves and their possible role in nerve impulse propagation Biophysical Reviews and Letters 05 7388.
\bibitem{tejocote16} Marat M. Rvachev and  Benjamin Drukarch. Surface Waves and Axoplasmic Pressure Waves in Action Potential Propagation: Fundamentally Different Physics or Two Sides of the Same Coin? Biophysical Reviews and Letters Vol. 20, No. 4 (2025) 283–289.       
\bibitem{tejocote17} Alexander Mengnjo, Alain M. Dikandé, Gideon A. Ngwa. Model of the nerve impulse with account of mechanosensory processes: Stationary solutions. J Appl Math Phys, 8 (2020), pp. 2091-2102.
\bibitem{tejocote18} Alexander Mengnjo, Jake Leonard Nkeck. On the hybrid model of nerve pulse: Mathematical analysis and numerical results. J Appl Math Phys, 11 (2023), pp. 2373-2396. 
\bibitem{tejocote19} R.A. El-Nabulsi. Emergence of lump-like solitonic waves in Heimburg–Jackson biomembranes and nerves fractal model. J R. Soc Interface, 19 (2022), Article 20220079. 
\bibitem{tejocote20} Laura R. González-Ramírez. Fractional-Order Traveling Wave Approximations for a Fractional-Order Neural Field Model. Front. Comput. Neurosci., 23 March 2022. 
\bibitem{tejocote020} Aly R. Seadawy, Asghar Ali, Ahmet Bekir. Solitary wave solutions of the nonlinear fractional soliton neuron model via application of five mathematical methods. Modern Phys Lett B (2025), Article 2550098 (20 pages).
\bibitem{tejocote21} Edgar Villagran Vargas, Andrei Ludu, Reinhold Hustert, Peter Gumrich, Andrew D. Jackson, Thomas Heimburg, Periodic solutions and refractory periods in the soliton theory for nerves and the locust femoral nerve, Biophysical Chemistry, Volume 153, Issues 2–3, 2011, Pages 159-167, ISSN 0301-4622,https://doi.org/10.1016/j.bpc.2010.11.001.
\bibitem{tejocote22} F. Contreras, F. Ongay, O. Pavón, M. Aguero. Non-topological solitons as travelling pulses along the nerve. Int J Mod Nonlinear Theory Appl, 02 (2013), Article 195200. 
\bibitem{tejocote23} O. Pavón-Torres, M.A. Agüero-Granados, M.E. Maguiña-Palma. Interaction and adiabatic evolution of orthodromic and antidromic impulses in the axoplasmic fluid. Phys Lett A, 521 (2024), Article 129740. 
\bibitem{tejocote24} Pavón-Torres, M.A. Agüero-Granados, Valencia-Torres. R. Adiabatic evolution of solitons embedded in lipid membranes. Phys Scr, 99 (2024), Article 125256. 
\bibitem{tejocote25} Rani, A.; Shakeel, M.; Kbiri Alaoui, M.; Zidan, A.M.; Shah, N.A.; Junsawang, P. Application of the $Exp-\varphi \xi$-Expansion Method to Find the Soliton Solutions in Biomembranes and Nerves. Mathematics 2022, 10, 3372. 
\bibitem{tejocote26} Razzaq, W., Akbulut, A., Zafar, A. et al. Solitary wave solutions of coupled nerve fibers model based on two analytical techniques. Opt Quant Electron 55, 591 (2023).
\bibitem{tejocote27} González-Gaxiola, O.; Biswas, A.; Moraru, L.; Alghamdi, A.A. Solitons in Neurosciences by the Laplace–Adomian Decomposition Scheme. Mathematics 2023, 11, 1080.
\bibitem{tejocote28} Shahzad T, Baber M Z, Qasim M, Sulaiman T A, Yasin M W and Ahmed N 2024 Explicit solitary wave profiles and stability analysis of biomembranes and nerves Modern Physics Letters B 38.
\bibitem{tejocote29} Tahira Jamal, Adil Jhangeer, Malik Zawwar Hussain. An anatomization of pulse solitons of nerve impulse model via phase portraits, chaos and sensitivity analysis. Chinese Journal of Physics 87 (2024) 496–509.
\bibitem{tejocote30} Ozsahin D U, Ceesay B, baber M Z, Ahmed N, Raza A, Rafiq M, Ahmad H, Awwad F A and Ismail E A A 2024 Multiwaves, breathers, lump and other solutions for the heimburg model in biomembranes and nerves Scientific Reports 14.
\bibitem{tejocote31} Younas, U., Muhammad, J., Almutairi, D.K. et al. Analyzing the neural wave structures in the field of neuroscience. Sci Rep 15, 7181 (2025).
\bibitem{tejocote32} Attia Rani, Muhammad Shakeel, Muhammad Sohail, Ibrahim Mahariq. The generalizing riccati equation mapping method's application for detecting soliton solutions in biomembranes and nerves, Partial Differential Equations in Applied Mathematics, Volume 15, 2025, 101300, ISSN 2666-8181,https://doi.org/10.1016/j.padiff.2025.101300.
\bibitem{tejocote33} C.S. Fedosejevs, \& M.F. Schneider, Sharp, localized phase transitions in single neuronal cells, Proc. Natl. Acad. Sci. U.S.A. 119 (8) e2117521119 (2022).
\bibitem{tejocote34} Jüri Engelbrecht, Kert Tamm, Tanel Peets. On mathematical modelling of solitary pulses in cylindrical biomembranes. Biomech Model Mechanobiol (2015) 14:159–167. 
\bibitem{tejocote35} Tanel Peets, Kert Tamm, Jüri Engelbrecht. On the role of nonlinearities in the Boussinesq-type wave equations. Wave Motion 71 (2017) 113–119. 
\bibitem{tejocote36} Jüri Engelbrecht, Kert Tamm \& Tanel Peets (2017) On solutions of a Boussinesq-type equation with displacement-dependent nonlinearities: the case of biomembranes, Philosophical Magazine, 97:12, 967-987. 
\bibitem{tejocote37} Tanel Peets, Kert Tamm, Päivo Simson, Jüri Engelbrecht. On solutions of a Boussinesq-type equation with displacement-dependent nonlinearity: A soliton doublet. Wave Motion 85(2019) 10–17. 
\bibitem{tejocote38} V.A. Mendoza-Millán, J.L. Larios-Ferrer, J. Samuel Millán, M.A. Agüero-Granados, D.M. Galván-Arellano, O. Pavón-Torres. Lambert W-Kink solitons arising from higher-order nonlinearities of lipid membranes. Chaos, Solitons \& Fractals, Volume 201, Part 2, 2025, 117260, ISSN 0960-0779, https://doi.org/10.1016/j.chaos.2025.117260.
\bibitem{tejocoteprepre39} Demirkaya, A., Decker, R., Kevrekidis, P.G. et al. Kink dynamics in a parametric $\phi^6$ system: a model with controllably many internal modes. J. High Energ. Phys. 2017, 71 (2017). https://doi.org/10.1007/JHEP12(2017)071.
\bibitem{tejocotepre39} Amado, A., Mohammadi, A. A $\phi^{6}$ soliton with a long-range tail . Eur. Phys. J. C 80, 576 (2020). https://doi.org/10.1140/epjc/s10052-020-8162-9
\bibitem{tejocote39} O. Cornejo-Pérez and H. C. Rosu. Nonlinear Second Order Ode’s -factorization and particular solutions- Progress of Theoretical Physics, Vol. 114, No. 3 (2005).
\bibitem{tejocote40} González, H.C. Rosu, O. Cornejo-Pérez, S.C. Mancas, Factorization conditions for nonlinear second-order differential equations, in: S. Manukure, W.-X. Ma (Eds.), Nonlinear and Modern Mathematical Physics-Proceedings 2022, Springer, USA, 2024, pp. 81–99. 
\bibitem{tejocote41} H. C. Rosu and O. Cornejo-Pérez. Supersymmetric pairing of kinks for polynomial nonlinearities. Phys. Rev. E 71, 046607 (2005).
\bibitem{tejocote42} J. A. Onana Inouga, S. E. Mkam Tchouobiap, M. Siewe Siewe, and F. M. Moukam Kakmeni. Action potential-like modes as modulated waves in an extended soliton model for biomembranes and nerves. AIP Advances 15, 015035 (2025). 
\bibitem{tejocote045} Kowalevski, S.: Sur le probleme de la rotation d’un corps solide autour d’un point fixe. Acta Math. 12(1), 177–232 (1889). https://doi.org/10.1007/BF02592182.
\bibitem{tejocote046} Kowalevski, S.: Sur une propriété du systém d’uations différentielles qui définit la rotation d’un corps solide autour d’un point fixe. Acta Math. 14(1), 81–93 (1890). https://doi.org/10.1007/BF02413316. 
\bibitem{tejocote0046} Nikolay A. Kudryashov. Painlev\'e Test, First Integrals and Exact Solutions of Nonlinear Dissipative Differential Equations. Regular and Chaotic Dynamics, 2025, Vol. 30, No. 5, pp. 819–836.
\bibitem{tejocote047} Abdul-Majid Wazwaz, Compactons, solitons and periodic solutions for some forms of nonlinear Klein–Gordon equations, Chaos, Solitons \& Fractals, Volume 28, Issue 4, 2006, Pages 1005-1013, ISSN 0960-0779,https://doi.org/10.1016/j.chaos.2005.08.145.
\bibitem{tejocote45} Norman Cruz, A. Hernández-Almada, Octavio Cornejo-Pérez. Constraining a causal dissipative cosmological model. Physical Review D 100, 083524 (2019).  
\bibitem{tejocote46} Belinchón, J.A., Cornejo-Pérez, O. \& Cruz, N. Exact solutions of a causal viscous FRW cosmology within the Israel–Stewart theory through factorization. Gen Relativ Gravit 54, 10 (2022). 
\bibitem{tejocote47} Dragana Rankovic, DraganPrekrat, Anna Batova, and Slobodan Zdravkovic. Stability of subsonic and supersonic solitons in DNA. Chaos 36, 013147 (2026); doi: 10.1063/5.0277901.
\bibitem{tejocote48} Stefan C. Mancas, Haret C. Rosu. Integrable dissipative nonlinear second order differential equations via factorizations and Abel equations. Physics Letters A 377 (2013) 1434–1438. 
\bibitem{tejocote51} Tamaghna Hazra, V. K. Chandrasekar, R. Gladwin Pradeep, and M. Lakshmanan. Exact solutions of coupled Liénard-type nonlinear systems using factorization technique. J. Math. Phys. 53, 023511 (2012); doi: 10.1063/1.3684956. 
\bibitem{tejocote52} Ajey K. Tiwari, S.N. Pandey, V.K. Chandrasekar, M. Lakshmanan. Factorization technique and isochronous condition for coupled quadratic and mixed Liénard-type nonlinear systems. Applied Mathematics and Computation 252 (2015) 457–472.
\bibitem{tejocote50} G.González, O.Cornejo-Pérez, J.de la Cruz, H. C.Rosu. Isochronous waveforms of Liénard equations via commutative factorization. Physics Letters A 564 (2025) 131087. 
\bibitem{mafer00} Yousef AbuHour, Mohammed Banikhalid and Amirah Azmi. Solutions of the generalized Heimburg–Jackson model for membrane pulses. Z. Angew. Math. Phys. (2026) 77:193.
\bibitem{phtrans} H. Yasuda, L.M.Korpas, and J.R.Raney. Transition Waves and Formation of Domain Walls in Multistable Mechanical Metamaterials. Physical Review Applied 13, 054067 (2020).
\bibitem{sexticpotential1} Escobar Ruiz, A.M., Mendoza Tavera, A.N., Sagar, R.P. et al. Wigner-entropy and negativity signatures of tunneling in a sextic double well. Eur. Phys. J. Plus 141, 240 (2026).
\bibitem{sexticpotential2} Mendoza Tavera, A.N., Escobar Ruiz, A.M. \& Sagar, R.P. Entropic Characterization of Tunneling and State Pairing in a Quasi-exactly Solvable Sextic Potential. Int J Theor Phys 64, 322 (2025).
\bibitem{sexticpotential3} Alonso Contreras-Astorga, Adrian M Escobar-Ruiz and Román Linares. The SUSY partners of the QES sextic potential revisited. Phys. Scr. 99 (2024) 025223.
\bibitem{susy1} E Cooper et al. Supersymmetry and quantum mechanics. Physics Reports 251 (1995) 267-385. 
\bibitem{tejocote43} R.J.H. Cloots, J.A.W. van Dommelen, M.G.D. Geers. A tissue-level anisotropic criterion for brain injury based on microstructural axonal deformation. Journal of the mechanical behavior of biomedical materials 5 (2012) 41-52. 
\bibitem{tejocote44} Delteil C, Manlius T, Bailly N, Godio-Raboutet Y, Piercecchi-Marti MD, Tuchtan L, Hak JF, Velly L, Simeone P, Thollon L. Traumatic axonal injury: Clinic, forensic and biomechanics perspectives. Leg Med (Tokyo). 2024 Sep;70:102465. doi: 10.1016/j.legalmed.2024.102465. Epub 2024 Jun 2. PMID: 38838409. 
\bibitem{feriha1} Arévalo, E., Gaididei, Y. \& Mertens, F. Soliton dynamics in damped and forced Boussinesq equations. Eur. Phys. J. B 27, 63–74 (2002). https://doi.org/10.1140/epjb/e20020130.
\bibitem{feriha2} Fan, Kai, Zhou, Cunlong, Exact Solutions of Damped Improved Boussinesq Equations by Extended (G'/G)-Expansion Method, Complexity, 2020, 4128249, 14 pages, 2020. https://doi.org/10.1155/2020/4128249. 
\bibitem{tejocote013} Guirland, C., Zheng, J.Q. (2007). Membrane Lipid Rafts and Their Role in Axon Guidance. In: Bagnard, D. (eds) Axon Growth and Guidance. Advances in Experimental Medicine and Biology, vol 621. Springer, New York, NY. 
\bibitem{tejocote0013} P.C. Bressloff, Waves in Neural Media (Springer, Berlin, 2014).
\bibitem{tejocote00013} Mareš, J.J., Špička, V. \& Hubík, P. On physical processes controlling nerve signalling. Eur. Phys. J. Spec. Top. 232, 3561–3576 (2023). 
\bibitem{llamaeterna0} Volkov, A.G., Foster, J.C., Ashby, T.A., Walker, R.K., Johnson, J.A. and Markin, V.S. (2010), Mimosa pudica: Electrical and mechanical stimulation of plant movements. Plant, Cell \& Environment, 33: 163-173.
\bibitem{llamaeterna1} Stolarz, M. \& Trebacz, K. (2021) Spontaneous rapid leaf movements and action potentials in Mimosa pudica L. Physiologia Plantarum, 173(4), 1882–1888. 
\end{thebibliography}
\end{document}